\documentclass[12pt]{article}

\usepackage[a4paper,top=2.5cm,left=1.75cm,right=1.75cm,bottom=3.5cm]{geometry}
\usepackage[utf8]{inputenc}
\usepackage[T1]{fontenc}
\usepackage{authblk}
\usepackage{amsmath,amssymb,amsfonts,amsthm}
\usepackage{lineno}
\usepackage[table,svgnames,dvipsnames]{xcolor}
\usepackage{graphicx}
\usepackage{tabularx}
\usepackage[colorlinks=true, linkcolor=black, urlcolor=blue, citecolor=blue]{hyperref}
\usepackage{booktabs}
\usepackage[labelfont=bf,margin={1.75cm,1.75cm},font=small]{caption}
\usepackage{rotating}
\usepackage{multirow}
\usepackage{titlesec}
\usepackage{siunitx}

\newcommand{\kb}{k_{\mathrm{B}}}

\title{Temperature transferable Machine Learned Coarse Grained model for proteins}

\author[1]{Jacopo Venturin}
\author[1,2,3]{Cecilia Clementi\textsuperscript{*}}

\affil[1]{Department of Physics, Freie Universit\"at Berlin, \emph{Arnimallee 12}, 14195, Berlin, Germany}
\affil[2]{Center for Theoretical Biological Physics, Rice University, Bioscience Research Collaborative, \emph{6500 Main Street}, Houston, 77005, TX, USA}
\affil[3]{Department of Chemistry, Rice University, \emph{6100 Main Street}, Houston, 77030, TX, USA}

\date{}

\begin{document}

\maketitle

\begin{center}
\textsuperscript{*}Corresponding authors. E-mail: \href{mailto:cecilia.clementi@fu-berlin.de}{cecilia.clementi@fu-berlin.de}\\
\end{center}

\abstract{Coarse-grained (CG) molecular simulations offer an efficient alternative to atomistic molecular dynamics to study large and complex biological systems. The accuracy of CG simulations has been increased dramatically by the introduction of machine-learned coarse-grained (MLCG) models. However, these models are typically designed to be used at a single thermodynamic point, lack temperature transferability, and can not be used to predict temperature dependent quantities like the heat capacity.
Here we introduce a thermodynamically informed, temperature-transferable MLCG framework for proteins that explicitly decomposes the CG potential of mean force (PMF) into its energetic and entropic components. The model architecture enforces an exact thermodynamic relation between the energetic and entropic components of the PMF and guarantees physically consistent extrapolation and interpolation across temperature regimes.
We validate this framework on an extensive dataset spanning a total of \SI{250}{\micro\second} of molecular dynamics simulations across five temperatures between \SI{300}{\kelvin} and \SI{400}{\kelvin} for the Chignolin protein, and demonstrate that it reproduces the temperature dependency of the reference atomistic free energy surfaces, correcting the temperature-unaware baselines. Furthermore, we show that it is possible to apply an inexpensive, post-hoc temperature-dependent correction that does not require retraining the MLCG potential, accurately recovering the atomistic heat capacity at different temperatures. Overall, this work provides a physically grounded pathway toward thermodynamically transferable MLCG simulations of complex biomolecular systems.}

\clearpage

\section{Introduction}
Coarse-grained (CG) modeling provides a powerful framework to extend the spatial and temporal scales inaccessible to molecular simulations at atomistic resolution by systematically reducing the number of degrees of freedom. Because of the reduced complexity, CG models enable the study of large-scale conformational changes and long-time dynamics that would otherwise be computationally prohibitive~\cite{clementi2008coarse,stevens2023molecular,charron2025navigating,noid2024rigorous}. One of the central challenges in CG modeling is the definition of effective interactions between the CG particles (or \textit{beads}) that can accurately describe the desired properties of a system of interest.

From a bottom-up perspective~\cite{noid2008multiscale, noid2013perspective, jin2022bottom}, the effective CG interaction potential that can reproduce the thermodynamics of a reference finer-grained system is given by the potential of mean force (PMF), resulting from the renormalization of the CG degrees of freedom. While the PMF provides a thermodynamically consistent description of the CG system, its direct evaluation is in practice intractable for complex molecular systems. To overcome this limitation, a variety of approximation strategies have been developed~\cite{shell2008relative,noid2013perspective,noid2024rigorous}, among which force-matching~\cite{noid2008multiscale} has emerged as one of the most widely used approaches.

The accuracy of CG potentials for molecular systems has been significantly improved by recent advances in machine learning~\cite{wang2019machine,durumeric2023machine}, particularly by using message-passing graph neural networks GNNs) to approximate the PMF~\cite{husic2020coarse, majewski2023machine, charron2025navigating,durumeric2026learning}. 
Neural network can learn complex many-body interactions directly from atomistic reference data~\cite{wang2019machine,wang2021multi}; however, they typically approximate the PMF only at a specific thermodynamic condition. This restricts their thermodynamic transferability and prevents their application across different state points, e.g. different temperatures.

From a thermodynamic perspective, in a canonical ensemble setting, the CG PMF corresponds to a Helmholtz free energy~\cite{dunn2016van}, and can be decomposed into energetic and entropic contributions~\cite{dunn2016van, kidder2021energetic, szukalo2021investigating}. This decomposition plays a crucial role in determining the temperature dependence of the PMF, which is generally dominated by the entropic contribution. Typically, machine-learned CG (MLCG) models implicitly encode both energy and entropy within a single effective potential and do not allow for an explicit separation of these contributions. As a consequence, such models cannot be used to predict quantities directly related to the energetic or entropic components of the PMF (e.g., the heat capacity), nor can they be consistently extrapolated across temperatures~\cite{mussi2025predicting}.

To overcome this limitation, previous work has suggested a linear approximation of the PMF temperature dependency~\cite{kidder2021energetic}. By estimating both the PMF and its energetic component at a single thermodynamic state, it is possible to determine the temperature derivative of the PMF and use it to linearly extrapolate the free energy to different temperatures. In practice, this can be achieved by fitting two independent models: one via force matching, approximating the full PMF, and the other via energy matching, approximating the energetic component of the PMF~\cite{kidder2021energetic, szukalo2021investigating}. This approach is theoretically robust but may be difficult to implement in practice, particularly when the solvent degrees of freedom are coarse-grained away (e.g., ~\cite{husic2020coarse, charron2025navigating}) because of the large energy variance over the  different fine-grained configurations consistent with a coarse-grained one, which makes the energy matching procedure extremely noisy.

Building on previous theoretical developments~\cite{kidder2021energetic,szukalo2021investigating}, here we introduce a temperature-dependent MLCG modeling framework, which explicitly decomposes the PMF into energetic and entropic components in a thermodynamically consistent manner. By enforcing an exact thermodynamic relation between energy and entropy, our GNN architecture, coupled with a joint force- and energy-matching loss, ensures a stable training process and yields a model that remains physically grounded across temperatures.

We use the MLCG modeling framework~\cite{husic2020coarse, charron2025navigating} to learn a flexible and accurate representation of the PMF. Simultaneously, we incorporate temperature-dependent, physically motivated energy priors and constraints to obtain a correct physical description in the neural network extrapolation. By jointly optimizing forces and energy, the proposed approach enables the recovery of thermodynamic properties on an extended range of temperatures rather than a single thermodynamic state.

We demonstrate the proposed modeling framework on a dataset specifically designed for this task, consisting of atomistic molecular dynamics (MD) simulations of the solvated chignolin protein at different temperatures. We show that we can correctly interpolate and extrapolate the system thermodynamic behavior and accurately reproduce the free energy landscapes and heat capacity of the atomistic model on a range of temperature between \SI{300}{\kelvin} and \SI{400}{\kelvin}. These results highlight the importance of incorporating thermodynamic consistency into MLCG models and open the way towards thermodynamically transferable and physically grounded CG simulations of complex systems.

\begin{figure}
    \centering
    \includegraphics[width=1.0\linewidth]{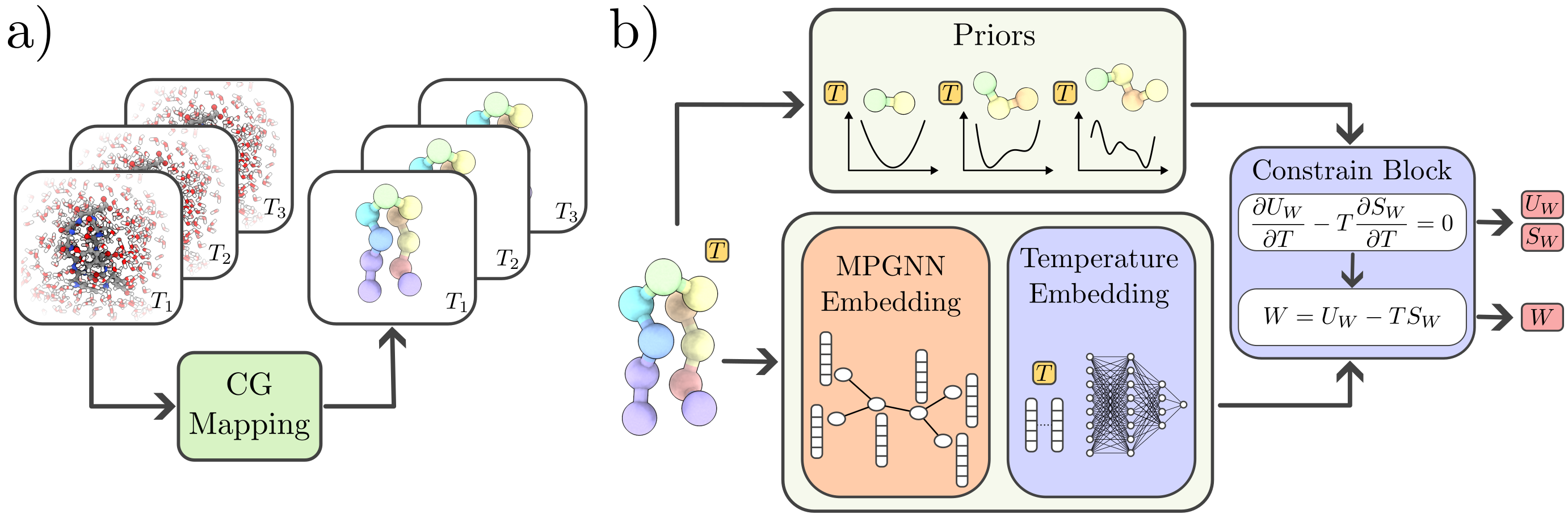}
    \caption{\textbf{Overview of the temperature-transferable MLCG framework.} \textbf{a)} Generation of the multi-temperature dataset by projecting reference atomistic data from multiple temperature points onto the CG resolution. \textbf{b)} Schematic representation of the model architecture, highlighting its extension over the MLCG framework through the integration of temperature-dependent prior models, a temperature embedding layer, and the constraint block enforcing the thermodynamic relation.}
    \label{fig:method}
\end{figure}

\section{Methods}
\subsection{Theory}
Let us consider a generic system of $n$ atoms with atomic positions $\mathbf{r}=\{\mathbf{r}_1, \dots, \mathbf{r}_n\}$ and a temperature-independent atomistic potential $u(\mathbf{r})$. In a canonical setting, with system volume $V$ and temperature $T$, the atomic configurations are distributed according to the Boltzmann distribution:
\begin{equation}
    \mathbb{P}_\mathbf{r}(\mathbf{r}) = \frac{e^{-\beta u(\mathbf{r})}}{\mathcal{Z}_\mathbf{r}}, \qquad \mathcal{Z}_\mathbf{r}= \int_{V^n} d\mathbf{r}\,e^{-\beta u(\mathbf{r})},
    \label{eq:BoltzAA}
\end{equation}
where $\beta = (\kb T)^{-1}$.

A bottom-up coarse-grained (CG) model provides a systematic framework for reducing the dimensionality of a molecular system while maintaining its structural and thermodynamic properties. Coarse-grained coordinates $\mathbf{R} = \{\mathbf{R}_1, \dots, \mathbf{R}_N\}$, with $N < n$, are typically obtained from the atomistic ones via a linear mapping operator $\Xi: \mathbb{R}^{3n} \mapsto \mathbb{R}^{3N}$ such that 
\begin{equation}
    \mathbf{R} = \Xi\,\mathbf{r}.
\end{equation}
Once the mapping operator is defined, one can define the many-body potential of mean force (PMF) $W$ such that 
\begin{equation}
    W(\mathbf{R}) = -k_B T \ln\left( V^{N-n}\int_{V^n} d\mathbf{r}\,e^{-\beta u(\mathbf{r})}\delta(\Xi\, \mathbf{r}-\mathbf{R})\right).
    \label{eq:w_therm_consistency}
\end{equation}
From a statistical mechanics perspective, the PMF is associated with the Boltzmann weight of the ensemble of atomistic configurations mapping to a coarse-grained configuration $\mathbf{R}$:
\begin{equation}
    \mathbb{P}_\mathbf{R}(\mathbf{R}) = \frac{e^{-\beta W(\mathbf{R})}}{\mathcal{Z}_\mathbf{R}}, \qquad \mathcal{Z}_\mathbf{R}= \int_{V^N} d\mathbf{R}\,e^{-\beta W(\mathbf{R})}.
    \label{eq:BoltzCG}
\end{equation}
While the direct evaluation of the integral in Eq.~\eqref{eq:w_therm_consistency} is intractable in practice due to the high dimensionality of the conformational phase space, various methods have been proposed to approximate the PMF from data. One of the most common is the force-matching method~\cite{noid2008multiscale}, where $W$ is obtained by minimizing the force-matching functional:
\begin{equation}
    \chi[U] = \langle \| \Xi_f \mathbf{f}(\mathbf{r}) + \nabla U(\Xi \mathbf{r}) \|^2 \rangle_\mathbf{r}.
\end{equation}
Here, $\mathbf{f}(\mathbf{r}) = -\nabla_\mathbf{r}u(\mathbf{r})$ represent the atomistic forces associated with the configuration $\mathbf{r}$, and $\Xi_f: \mathbb{R}^{3n} \mapsto \mathbb{R}^{3N}$ is the force-mapping operator, whose definition is related to the choice of $\Xi$~\cite{kramer2023statistically}.
Crucially, the global minimum of the force-matching functional in the $L_2$ functional space corresponds to the potential of mean force, Eq.~(\ref{eq:w_therm_consistency}), at the thermodynamic state associated with the canonical ensemble average $\langle \cdot \rangle_\mathbf{r}$.

\subsubsection{Energetic and entropic decomposition}

As discussed extensively in previous work~\cite{dunn2016van, kidder2021energetic, szukalo2021investigating}, the PMF can be interpreted as a configuration-dependent Helmholtz free energy. It can thus be decomposed into energetic and entropic contributions:
\begin{equation}
    \label{eq:energy_entropy}
    W(\mathbf{R}) = U_W(\mathbf{R}) - T S_W(\mathbf{R}),
\end{equation}
where the dependence on the coarse-grained configuration $\mathbf{R}$ is shown explicitly, while other thermodynamic variables are kept implicit for conciseness.

The energetic contribution, $U_W(\mathbf{R})$, is defined as the average atomistic potential energy for a given coarse-grained configuration:
\begin{equation}
    \label{eq:def_Uw}
    U_W(\mathbf{R}) = \langle u(\mathbf{r}) \rangle_{\mathbf{r}|\mathbf{R}},
\end{equation}
where the conditional ensemble average is performed according to the conditional distribution:
\begin{equation}
    \mathbb{P}_{\mathbf{r}|\mathbf{R}} = \frac{\mathbb{P}_\mathbf{r}(\mathbf{r})\delta(\Xi\,\mathbf{r}-\mathbf{R})}{\mathbb{P}_\mathbf{R}(\mathbf{R})}
\end{equation}
and the Boltzmann distributions $\mathbb{P}_\mathbf{r}(\mathbf{r})$ and $\mathbb{P}_\mathbf{R}(\mathbf{R})$ are defined by Eq.~(\ref{eq:BoltzAA}) and  (\ref{eq:BoltzCG}), respectively.
Given Eqs.~\eqref{eq:energy_entropy} and \eqref{eq:def_Uw}, the entropic contribution can be expressed as:
\begin{equation}
\label{eq:def_Sw}
\begin{aligned}
    S_W(\mathbf{R}) &= k_B \langle \beta u(\mathbf{r}) \rangle_{\mathbf{r}|\mathbf{R}} - k_B \beta W(\mathbf{R}) \\
    &= k_B \langle \beta u(\mathbf{r}) - \beta W(\mathbf{R}) \rangle_{\mathbf{r}|\mathbf{R}} \\
    &= -k_B \left\langle \ln \left[ \frac{V^N \mathbb{P}_{\mathbf{R}}(\mathbf{R})}{V^n \mathbb{P}_{\mathbf{r}}(\mathbf{r})} \right] \right\rangle_{\mathbf{r}|\mathbf{R}},
\end{aligned}
\end{equation}
where we have utilized the fact that $\langle W(\mathbf{R}) \rangle_{\mathbf{r}|\mathbf{R}} = W(\mathbf{R})$.

The role of $S_W$ becomes evident when considering the total differential of the PMF:
\begin{equation}
    \label{eq:differential_W}
    dW = -\langle \Xi_f \mathbf{f} \rangle_{\mathbf{r}|\mathbf{R}} \cdot d\mathbf{R} - \langle P(\mathbf{r}) \rangle_{\mathbf{r}|\mathbf{R}} dV - S_W dT,
\end{equation}
where $P(\mathbf{r})$ is the instantaneous atomistic pressure. The temperature dependence of the PMF is governed by $S_W(\mathbf{R})$ through the relation
\begin{equation}
    S_W(\mathbf{R}) = -\frac{\partial W(\mathbf{R})}{\partial T},
\end{equation}
that remains valid even when $U_W$ and $S_W$ are themselves temperature-dependent.  

It is important to highlight that, since from the definition in Eq.~\eqref{eq:def_Sw} $S_W<0$, the PMF is always lower bounded from the purely energetic contribution $U_W(\mathbf{R})$. Consequently, $W$ cannot be used directly to compute energetic properties of the system. A primary example is the constant-volume specific heat:
\begin{equation}
    c_V = \frac{\partial}{\partial T}\langle u \rangle_\mathbf{r} = \frac{\partial}{\partial T}\langle U_W \rangle_\mathbf{R},
\end{equation}
which underscores the necessity of maintaining an explicit and correct representation of the energetic component within a coarse-grained model.

\subsubsection{GNN for learning MLCG interaction potential}

As extensively discussed in previous work~\cite{wang2019machine,husic2020coarse, charron2025navigating}, the minimization of the force-matching functional for complex systems requires flexible and expressive multi-body functional forms. This leads naturally to the use of neural network, and in particular, message-passing graph neural networks (GNNs), to model the PMF.

In general terms, in MLCG models a GNN computes the total molecular energy as a sum of atomic contributions. These are obtained by mapping learned local atomic embeddings to scalar values via a readout function. The embedding vectors are generated through a message-passing scheme: each atom (or bead) $i$ is initialized with an embedding $\mathbf{h}_i^{(0)}$ based on inherent atomic properties (e.g., atomic number or bead type). These embeddings are then sequentially updated for a total of $\mathcal{T}$ iterations via:
\begin{equation}
    \mathbf{h}_i^{(t+1)} = f_t\left(\mathbf{h}_i^{(t)}, \mathbf{m}_i^{(t)}\right),
    \label{eq:update_step}
\end{equation}
where $f_t$ is a learned differentiable function and $\mathbf{m}_i^{(t)}$ is the aggregation of messages from the neighborhood $\mathcal{N}(i)$, defined by all atoms within a specified cutoff distance from the target atom $i$:
\begin{equation}
    \mathbf{m}_i^{(t)} = \bigoplus_{j \in \mathcal{N}(i)} g_t\left(\mathbf{h}_i^{(t)}, \mathbf{h}_j^{(t)}, \mathbf{e}_{ij}^{(t)}\right).
    \label{eq:message_step}
\end{equation}
Here, $g_t$ is a learned differentiable function, $\bigoplus$ denotes a permutation-invariant pooling operation (such as a sum or mean), and $\mathbf{e}_{ij}^{(t)}$ is an edge vector encoding structural information, e.g. interatomic displacements or distances.

The iterative application of Eq.~\eqref{eq:message_step} and \eqref{eq:update_step} yields a final representation for each atom or CG bead that incorporates information from its local receptive field. The atomic embeddings are then used to compute the total energy through a readout function $\mathcal{R}$:
\begin{equation}
    E = \sum_i \mathcal{R}\left(\left\{\mathbf{h}_i^{(t)}\right\}_{t=0}^{\mathcal{T}}\right).
\end{equation}

\subsubsection{Thermodynamic constraint on the GNN output}

An important relation between the energetic and entropic contributions can be obtained by considering the temperature derivative of Eq.~\eqref{eq:energy_entropy}:
\begin{equation}
    \frac{\partial W}{\partial T} = - S_W + \left(\frac{\partial U_W}{\partial T} - T \frac{\partial S_W}{\partial T}\right).
\end{equation}
Since the definition of entropy requires $\frac{\partial W}{\partial T} = -S_W$, it follows that the relation
\begin{equation}
    \label{eq:maxwell_relation}
    \frac{\partial U_W}{\partial T} - T \frac{\partial S_W}{\partial T} = 0
\end{equation}
must hold for every configuration $\mathbf{R}$. This is an extension of the well-known relation between internal energy and entropy, reformulated here for configuration-dependent quantities.

When utilizing a neural network potential to model the PMF, if an explicit representation of $U_W$ is also desired, the model must be designed such that the constraint in Eq.~\eqref{eq:maxwell_relation} is satisfied. A naive approach would be to introduce a penalty term during training (i.e., a soft constraint in the loss function). However, this does not guarantee that the constraint is exactly satisfied, potentially limiting the reliability of the model, particularly during extrapolation. 

A more robust approach is to design the neural network architecture to enforce the constraint by construction,  by using a constrained neural network framework~\cite{hendriks2020linearly}. In this framework, a generic unconstrained model $f(\mathbf{R})$ is forced to satisfy a constraint operator $\mathcal{G}$ by defining a transformation $\mathcal{M}$ such that the composition $\mathcal{G}\mathcal{M}$ is the null operator. 

In the specific case of the thermodynamic constraint of Eq.~(\ref{eq:maxwell_relation}), the linear operator $\mathcal{G}: \mathbb{R}^2 \mapsto \mathbb{R}$ can be defined by its action:
\begin{equation*}
    \mathcal{G}_T \begin{bmatrix} U_W \\ S_W \end{bmatrix} = \begin{bmatrix} \partial_T & -T\partial_T \end{bmatrix} \begin{bmatrix} U_W \\ S_W \end{bmatrix}, \quad \text{or in terms of } \beta: \quad \mathcal{G}_\beta \begin{bmatrix} U_W \\ S_W \end{bmatrix} = \begin{bmatrix} \partial_\beta & -\frac{1}{k_B \beta} \partial_\beta \end{bmatrix} \begin{bmatrix} U_W \\ S_W \end{bmatrix}.
\end{equation*}
The most natural design choices for the unconstrained network, though not unique, involve predicting either one or two scalar outputs. The latter directly aligns with predicting separate unconstrained energetic and entropic contributions to the PMF, which are subsequently bound by the constraint block. In practice, however, utilizing a single scalar output leads to a simpler model implementation and yields the elegant property that the unconstrained network's direct prediction corresponds to the PMF itself.
In fact, if we consider the output of the unconstrained model to be a scalar, and thus define the mapping $\mathcal{M}: \mathbb{R} \mapsto \mathbb{R}^2$. A possible solution to the equation $\mathcal{G}\mathcal{M} = \mathbb{O}$ is:
\begin{equation*}
    \mathcal{M}_T = \begin{bmatrix} 1 - T\partial_T \\ -\partial_T \end{bmatrix}, \quad \text{or in terms of } \beta: \quad \mathcal{M}_\beta = \begin{bmatrix} 1 + \beta\partial_\beta \\ k_B \beta^2 \partial_\beta \end{bmatrix}.
\end{equation*}

To train a CG model capable of simultaneously inferring the correct PMF and its energetic and entropic components, we optimize the model parameters to minimize a multi-objective loss function. This loss combines the standard force-matching (FM) objective with an energy-matching (EM) term~\cite{kidder2021energetic}:
\begin{equation}
    \mathcal{L} = \mathcal{L}_{FM} + \alpha \mathcal{L}_{EM}, 
    \label{eq:global_loss}
\end{equation}
where the individual components are defined as:
\begin{equation}
    \begin{cases}
        \mathcal{L}_{FM} &= \frac{1}{3N} \sum_{i=1}^{M} \| \Xi_f \mathbf{f}(\mathbf{r}_i) + \nabla_{\mathbf{R}} \hat{W}(\Xi\mathbf{r}_i; \boldsymbol{\theta}) \|^2 \\
        \mathcal{L}_{EM} &= \frac{1}{M} \sum_{i=1}^{M} | u(\mathbf{r}_i) - \hat{U}_W(\Xi\mathbf{r}_i; \boldsymbol{\theta}) |^2
    \end{cases}
\end{equation}
Here, $\alpha \in \mathbb{R}^+$ is a scaling parameter used to balance the numerical contributions of the two terms, $M$ is the number of samples in the training set, and $\hat{W}$ and $\hat{U}_W$ are the approximation of the PMF and its energetic contribution as estimated by the GNN model, parameterized by the set of weights $\boldsymbol{\theta}$.

\subsection{Dataset}

To train and validate our framework, we generated an extensive multi-temperature molecular dynamics (MD) dataset of the 10-residue mini-protein Chignolin (CLN025), totaling $\sim$\SI{250}{\micro\second} of aggregated simulation time. The system was simulated using the \texttt{OpenMM} engine~\cite{eastman2023openmm} at five distinct temperatures—\SI{300}{\kelvin}, \SI{320}{\kelvin}, \SI{350}{\kelvin}, \SI{380}{\kelvin} and \SI{400}{\kelvin}—with a total of $\sim$\SI{50}{\micro\second} collected at each temperature to ensure thorough sampling of the conformational landscape across various thermal regimes. For the CG representation, we employed a $C_\alpha$ mapping scheme that reduces each amino acid residue to a single representative bead centered at its $\alpha$-carbon atom, systematically integrating out all side-chain atoms and the explicit solvent environment. System coordinates, forces, and relevant thermodynamic observables (such as potential energy) were recorded every \SI{2}{\pico\second}. A full overview of the stored data, together with all technical details regarding the MD simulations are provided in the Supplementary Information.

\subsection{Temperature dependent prior energy terms}

To ensure the neural network potential remains in a physically meaningful region of the conformational space during MD simulations, preventing issues such as atomic overlaps or unphysical bond stretching/compressions, we follow previous work~\cite{wang2019machine,husic2020coarse,charron2025navigating} and incorporate prior energy terms. These prior terms, typically derived via Boltzmann inversion, are also temperature dependent and need to be incorporated into the framework of a  temperature-transferable MLCG model.

While determining the exact temperature dependence for complex learned distributions is non-trivial, it can be derived analytically for features following a Gaussian distribution, such as bonded distances. For a feature $x(\mathbf{R})$ distributed as $\mathbb{P}_x \propto e^{-\beta k (x-x_0)^2}$, the associated free energy $A(x) = -k_B T \ln \mathbb{P}(x)$ includes a logarithmic temperature term arising from the partition function $\mathcal{Z} \propto \sqrt{T}$ normalizing the Boltzmann probability. Within the temperature range typically considered for molecular systems, this dependence can be linearized, yielding a prior of the form
\begin{equation}
    A(x)\simeq k(x-x_0)^2 + a T + A_0.
\end{equation}

In this work, we consider temperature-dependent priors for bonds, angles, and dihedral angles, while a repulsion prior modeling the excluded volume of the CG beads is considered temperature-independent, consistent with previous works~\cite{ruza2020temperature,shinkle2024thermodynamic}. Further details regarding the specific functional forms and the fitting of these priors are provided in the Supplementary Information.

\begin{figure}
    \centering
    \includegraphics[width=1.\linewidth]{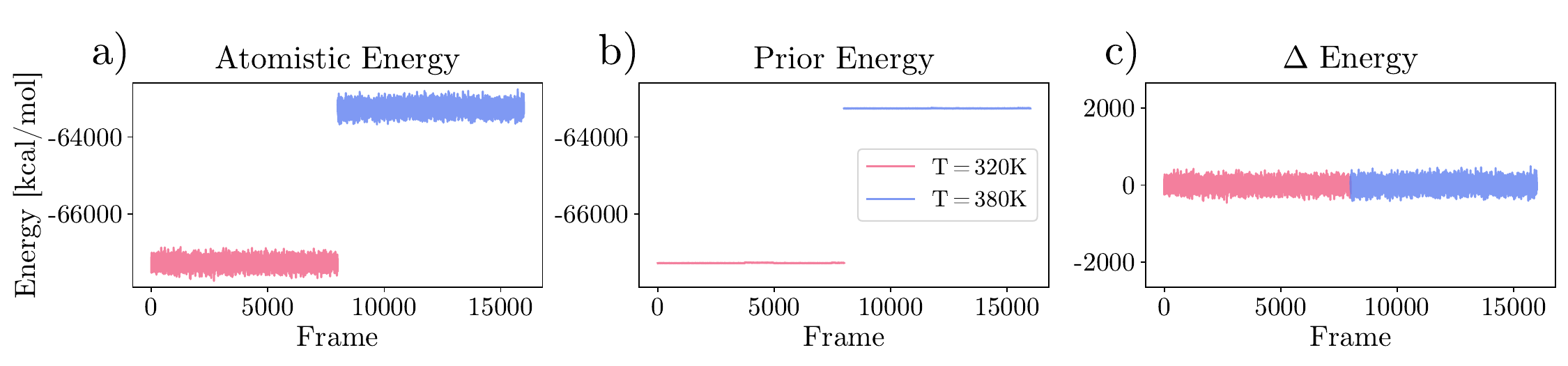}
    \caption{\textbf{Prior effect on $\Delta$-energy calculation.} \textbf{a)} Atomistic reference energy and \textbf{b)} prior energy profiles used to compute the \textbf{c)} resulting $\Delta$-energies for the two training temperatures (\SI{320}{\kelvin} and \SI{380}{\kelvin}). Applying the temperature-dependent energy shift rescales the $\Delta$-energies, thereby simplifying the neural network's learning procedure.}
    \label{fig:delta_energy}
\end{figure}

Importantly, to account for the mean energy shift, we introduce an extra prior term dependent solely on temperature. This term is crucial when energy matching is performed, as it shift the atomistic energy fluctuations used during training (see Fig.~\ref{fig:delta_energy}), simplifying the learning procedure. 
If we use training data at two different temperature values, we can at most fit a linear energetic shift induced by the temperature, $U_W^{\text{shift}} = aT + b$, and the thermodynamic relation of Eq.~\eqref{eq:maxwell_relation} dictates the functional form of the contribution to the PMF:
\begin{equation}
    W^{\text{shift}} = aT(1 - \ln T) + b.
\end{equation}
 as it is discussed below for the calculation of the heat capacity.

This temperature-dependent contribution does not affect the MD sampling forces, as it is spatially uniform; however, it is essential for the accurate estimation of thermodynamic observables that depend on the potential energy of sampled configurations. 
If training data at more than two temperatures are available, higher order polynomial functions can be used for a more accurate fit of the energetic shift with temperature. For instance, the calculation of heat capacity is highly sensitive to the functional form of the temperature shift; an incorrect assumption can lead to significant errors during temperature extrapolation. This issue is readily addressed since the calculation of the mean energy for a given temperature can be efficiently estimated from short atomistic MD allowing for a post-hoc correction of $U^{\text{shift}}$ and $W^{\text{shift}}$ via a simple scalar fit, bypassing the need to retrain the model.

\section{Results and Discussion}

To evaluate the performance of the proposed framework, we utilize the Chignolin dataset described above, comparing the temperature-dependent model against standard single-temperature models. The latter were trained at single state points (\SI{320}{\kelvin} and \SI{380}{\kelvin}) without explicit temperature dependency. In contrast, the temperature-dependent model was trained using a combined dataset from both temperatures (\SI{320}{\kelvin} and \SI{380}{\kelvin}), utilizing approximately half of the available trajectories from each to maintain a consistent total training data volume. The prior models for bonds, angles, and dihedrals were similarly parameterized using the \SI{320}{\kelvin} and \SI{380}{\kelvin} datasets, whereas the repulsion prior was fitted exclusively on the \SI{380}{\kelvin} data to accurately capture the short-range repulsive regions probed at higher thermal energies. 

The parameters for the linear global energy shift, $U_W^{\text{shift}}(T) = aT + b$, of the temperature-dependent model were fitted to the mean atomistic potential energies of the training ensembles at 320K and 380K. The introduction of this prior term is fundamental for the numerical stabilization of the model during training. By accounting for the baseline energy shift across temperatures, it prevents the GNN from attempting to compensate for large offsets in mean energy, thereby focusing the learning process on the relevant conformational fluctuations.

\subsection{Conformational sampling and thermodynamic consistency}

A central requirement for a temperature-transferable MLCG model is the ability to accurately reproduce the free energy profile of the atomistic reference across varying thermal regimes, maintaining the correct equilibrium distribution between folded and unfolded states. To evaluate the performance of our approach on this task, we performed molecular dynamics simulations at multiple temperatures using the differently trained MLCG models (temperature-dependent or single-temperature). A free energy profile as a function of the first Time-lagged Independent Component (TIC)~\cite{perez2013identification} of the atomistic data is then constructed for each model at every temperature. While the reference atomistic profiles are obtained via Markov State Model (MSM) reweighting~\cite{prinz2011markov, husic2018markov, hoffmann2021deeptime}, the CG simulations are already well-equilibrated and can be used directly to evalueate the free energy profile, as detailed in the Supplementary Information.

As reported in Fig.~\ref{fig:tics_results}, the single-temperature baselines fail to generalize at different temperatures: they are unable to maintain the correct balance between the folded and unfolded states at temperatures deviating from their training conditions. For example, the \SI{320}{\kelvin} model significantly over-stabilizes the unfolded state when tested at higher temperatures. In contrast, the temperature-dependent model accurately recovers the relative balance between folded unfolded states across the entire temperature range. Remarkably, this is valid not only for interpolation (\SI{350}{\kelvin}) but also for extrapolation to \SI{300}{\kelvin} and \SI{400}{\kelvin}.

We observe that, the transition region between folded and unfolded states is not well captured by any of the coarse-grained models, reflecting a general difficulty in accurately resolving free-energy barriers within force matching based approaches. This well known limitation arises because the force matching procedure primarily optimizes the agreement of mean forces in thermodynamically populated regions, thereby favoring accuracy in the metastable minima of the PMF rather than in rarely visited transition states~\cite{thaler2022deep,noid2024rigorous}.

\begin{figure}
    \centering
    \includegraphics[width=1.\linewidth]{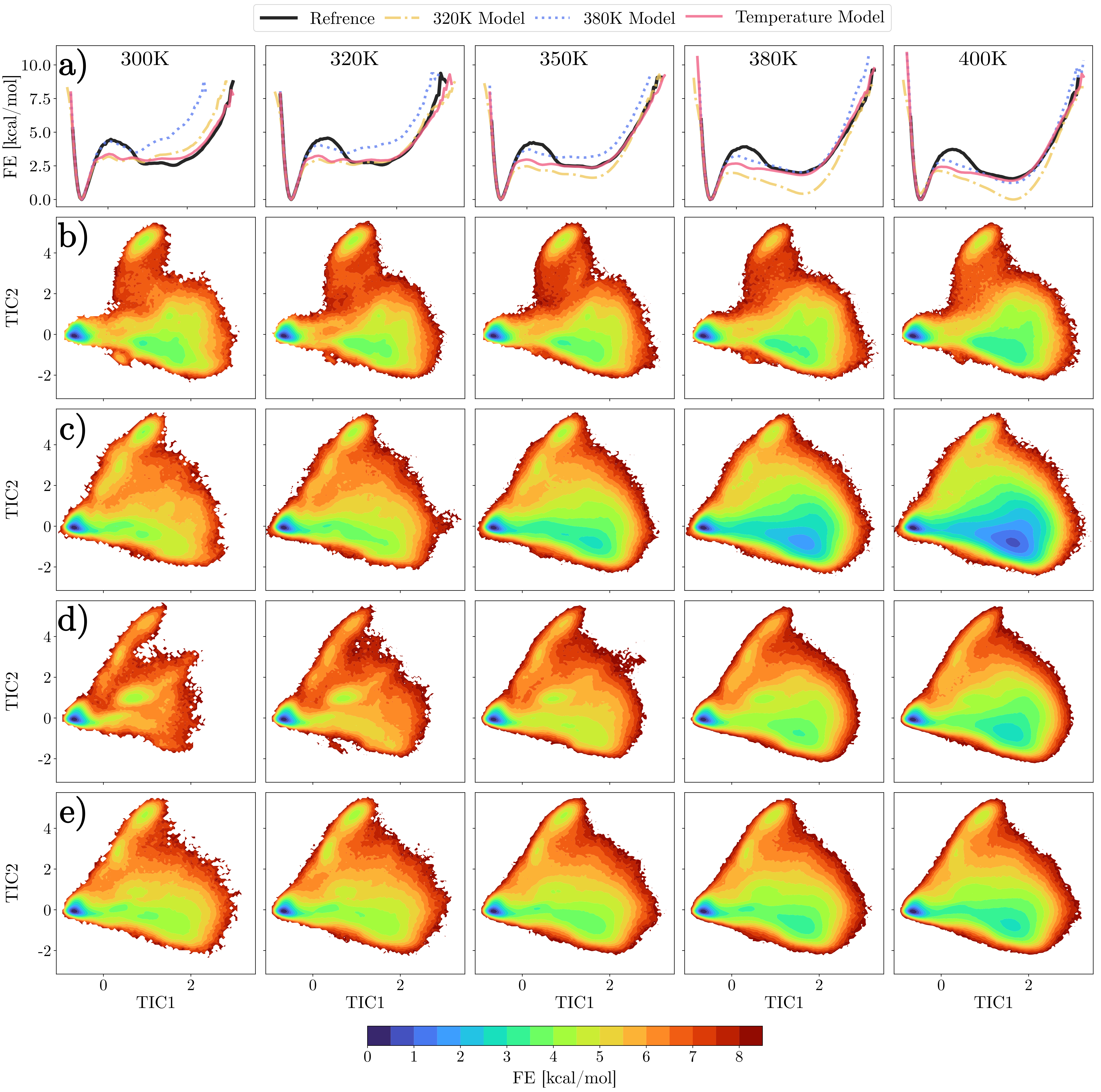}
    \caption{\textbf{Free energy prediction across different temperatures.} Free energy predicted by different models is compared with the atomistic reference, projected in TICA space. The two-dimensional projections highlight the presence of three main metastable states: folded (bottom left), unfolded (bottom right), and misfolded (top). \textbf{(a)} Projection of the free energy along the first TICA component. The two-dimensional projections correspond to \textbf{(b)} reference, \textbf{(c)} single-temperature model trained at \SI{320}{\kelvin}, \textbf{(d)} a single-temperature model trained at \SI{380}{\kelvin}, and \textbf{(e)} temperature dependent model.}
    \label{fig:tics_results}
\end{figure}

\begin{figure}
    \centering
    \includegraphics[width=0.8\linewidth]{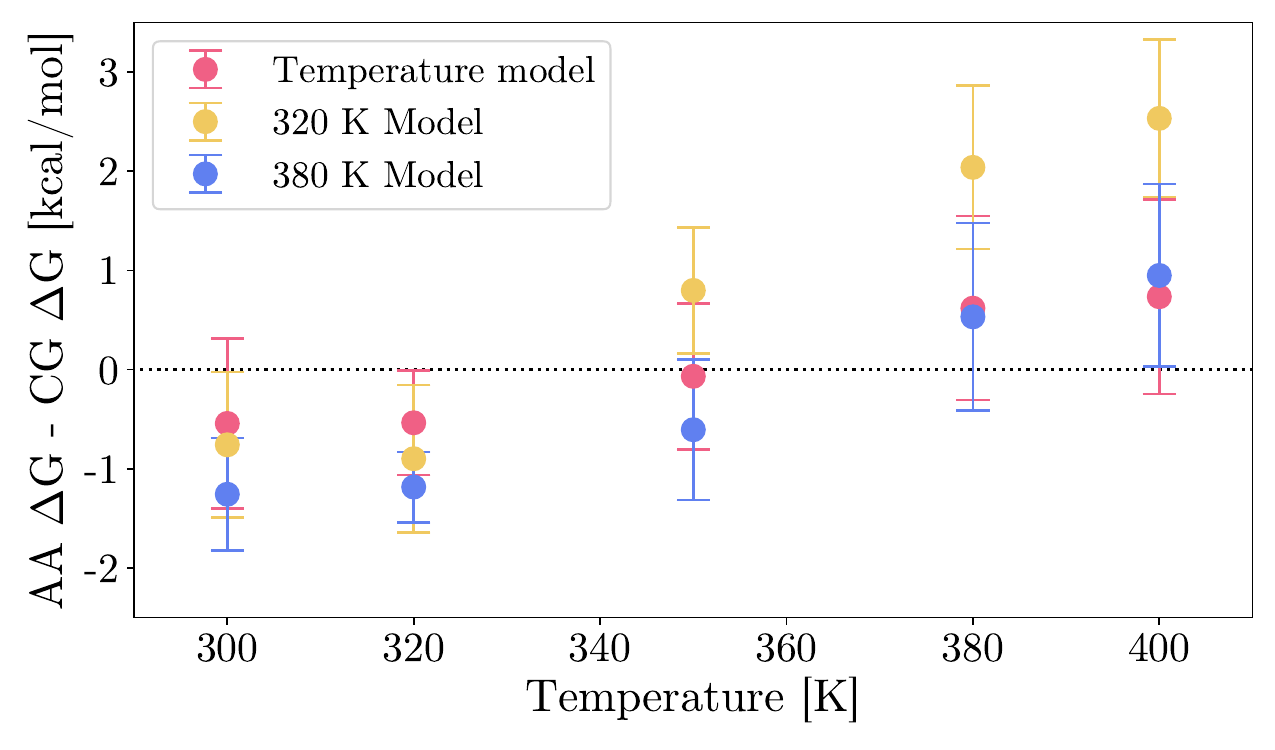}
    \caption{\textbf{Difference in $\Delta G$ between the atomistic reference and CG models.} Difference between the atomistic reference CG predicted free-energy, $\Delta G = -k_B T \ln(\mathbb{P}_{\text{folded}}/\mathbb{P}_{\text{unfolded}})$. Results for different CG models trained at a single temperature (\SI{320}{\kelvin} and \SI{380}{\kelvin}), as well as by the temperature-dependent CG model, across the investigated temperature range, are reported.}
    \label{fig:delta}
\end{figure}

To quantify this behavior, we estimate the relative stability between folded and unfolded states by computing the free energy difference, $\Delta G = -k_B T \ln(\mathbb{P}_{\text{folded}}/\mathbb{P}_{\text{unfolded}})$. The probabilities of the folded ($\mathbb{P}_{\text{folded}}$) and unfolded ($\mathbb{P}_{\text{unfolded}}$ including both misfolded and unfolded configurations) states are determined using a Perron Cluster Cluster Analysis (PCCA+)~\cite{roblitz2013fuzzy} on MSMs estimated from the respective molecular dynamics trajectories. Detailed procedures for the MSM construction and PCCA+ analysis are provided in the Supplementary Information.

The results of this analysis, shown in Fig.~\ref{fig:delta}, confirm that models trained at a single temperature fail to capture the correct thermodynamic trends. Conversely, the temperature-dependent model provides a good agreement with the atomistic $\Delta G$ for both interpolation (\SI{350}{\kelvin}) and extrapolation (\SI{300}{\kelvin} and \SI{400}{\kelvin}). This demonstrates that by explicitly accounting for the temperature dependence of the PMF, the model is able to recover the thermodynamic consistency with the reference AA system in the range temperature between \SI{300}{\kelvin} and \SI{400}{\kelvin}.

\subsection{Thermodynamic Observables: Heat Capacity}

An advantage of the proposed framework is its ability to predict thermodynamic quantities that depend directly on the mean internal energy, such as the isochoric heat capacity, $c_V$. Standard CG models, which implicitly bundle energetic and entropic components into a single potential, are fundamentally incapable of predicting such observables consistently across temperatures. 

\begin{figure}
    \centering
    \includegraphics[width=0.7\linewidth]{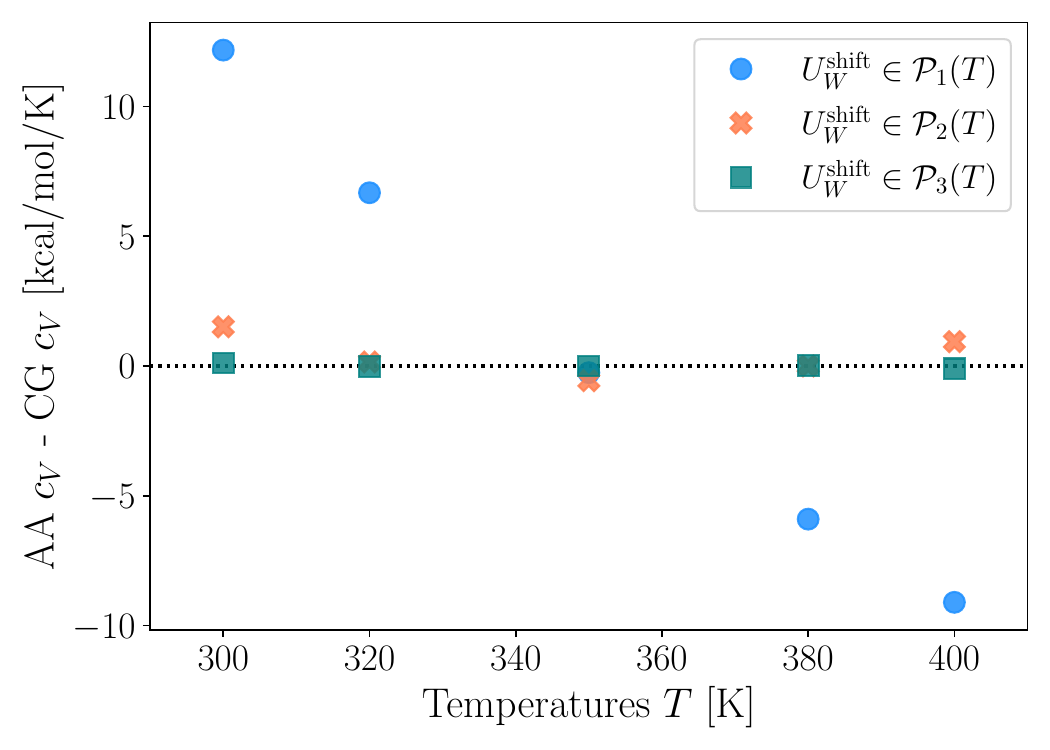}
    \caption{\textbf{Results for heat capacity prediction.} Absolute difference between the atomistic (AA) and coarse-grained (CG) heat capacities predicted by the temperature-dependent MLCG model. The linear energy shifts ($U_W^\text{shift}$) are modeled using polynomials in $T$ of first ($\mathcal{P}_1$), second ($\mathcal{P}_2$), and third ($\mathcal{P}_3$) order. Error bars are smaller than the marker size and are omitted for visual clarity.}
    \label{fig:cv}
\end{figure}

In Fig.~\ref{fig:cv}, we compare the reference heat capacity with the one predicted by the temperature-dependent model using different expressions for modeling the energy shift. It is immediately clear that the linear version of $U_W^{\text{shift}}$, while adequate for stabilizing the neural network during training, is insufficient for accurately evaluating the heat capacity outside the training states. 

As previously noted, this limitation can be readily addressed without retraining the potential. By leveraging the modularity of the framework, we applied a post-hoc scalar correction to $U_w^{\text{shift}}$ and $W^{\text{shift}}$ based on the mean energy of the system at additional temperature points. To demonstrate the efficiency of this approach, we estimated the mean potential energies at temperatures different from those used in training, utilizing only 250 MD frames per temperature. These estimates were then used to fit second- and third-order polynomial corrections for the energy shift; the detailed results for these additional $U_W^{\text{shift}}$ terms are reported in the Supporting Information. As shown in Fig.~\ref{fig:cv}, the inclusion of these higher-order terms provides an accurate agreement with the atomistic reference data. 

This result highlights that the structural features learned by the GNN are decoupled from the global energetic baseline, allowing the model to be refined for high-precision thermodynamic predictions through a simple, data-efficient scalar adjustment.

\section{Conclusions}
In this work, we have introduced a thermodynamically informed, temperature transferable MLCG model for proteins. By comparing our approach with standard, state of the art MLCG protein force fields, which are typically trained at a single temperature, we demonstrate that accounting for temperature dependence in the PMF is essential for accurately extrapolating and interpolating across different thermodynamic states of a protein system.

To validate the robustness of our model, we introduced a Chignolin dataset that spans a total of $\sim$~\SI{250}{\micro\second} of MD simulations across five different temperatures between \SI{300}{\kelvin} and \SI{400}{\kelvin}.

Leveraging the potential of mean force (PMF) energetic-entropic decomposition proposed in previous studies~\cite{kidder2021energetic}, our machine-learning potential satisfies the correct thermodynamic structure of the PMF by design: the physical consistency of the model is governed by the thermodynamic relation connecting the energetic and entropic components of the PMF, which we enforced as a constraint directly in the network architecture. Because this constraint is embedded by construction rather than learned through supplementary regularization terms in the loss function, our approach inherently guarantees the correct thermodynamic and physical structure of the PMF's energetic and entropic contributions when generalizing outside of the training domain, i.e. during extrapolation or interpolation.

The value of explicitly decomposing the PMF into its energetic and entropic components is further highlighted by the calculation of the isochoric heat capacity ($c_V$), a property fundamentally inaccessible to temperature-independent models. We found excellent agreement between the $c_V$ obtained via our iterative correction procedure and the atomistic reference, indicating that the heat capacity is primarily governed by a structure-independent energetic shift. 
This suggests that the dominant contribution to the system's heat capacity arises directly from the renormalized degrees of freedom and can be mostly decoupled from the energetic component of the PMF, i.e., the one contributing to the gradients with respect to the CG positions.

Importantly, this decoupling does not imply that the temperature dependence of the energetic and entropic terms in the PMF can be neglected, as demonstrated by the shifting balance between the folded and unfolded states on the free energy surface, which cannot be correctly recovered without the explicit temperature encoding of the force field.

\section*{Acknowledgments}
We thank Frank No\'e and members of the Clementi's group for insightful discussions and comments on the manuscript. 
We gratefully acknowledge funding from the Deutsche Forschungsgemeinschaft DFG (SFB/TRR 186, project A12; SFB 1114, projects B03, B08, and A04), the National Science Foundation (PHY-2019745), the Einstein Foundation Berlin (project 0420815101), the Bundesministerium f\"ur Bildung und Forschung BMBF (project FAIME 01IS24076), and computing time provided on the supercomputer Lise at NHR@ZIB as part of the NHR infrastructure (project beb00040). The authors also gratefully acknowledge the Gauss Centre for Supercomputing e.V. (www.gauss-centre.eu) for funding this project by providing computing time on the GCS Supercomputer JUWELS at J\"ulich Supercomputing Centre (JSC) (project mlcg).
We thank the HPC Services of FUB-IT and the Physics department of Freie Universit\"at Berlin.

\section*{Data availability}
The dataset used to train and validate the models and the configuration files for training and simulation will be available upon publication.

\section*{Code availability}
The codebase used for this work will be made available upon publication.

\newpage
\section*{Supplementary Information}

\setcounter{table}{0}
\setcounter{figure}{0}
\setcounter{section}{0}
\vspace{0.5cm}

\section{Prior models definition}

As discussed in the main text, a careful definition of prior energy terms is fundamental to prevent the model from sampling unphysical configurations during Molecular Dynamics (MD) simulations. In the context of a temperature-dependent machine-learned coarse-grained (MLCG) model, further care must be taken to design models capable of properly tracing the potential of mean force (PMF) behavior across the temperature interval of interest. 

Following previous work~\cite{husic2020coarse,charron2025navigating}, our model includes prior energy terms for bonds, angles, and dihedrals between two, three and four consecutive beads in the protein chain, respectively. All prior energy terms were fitted on the \SI{320}{\kelvin} and \SI{380}{\kelvin} datasets. To keep a concise notation, we omit to note the explicit dependency of the prior energy terms on the CG bead type (i.e., the residue type associated with the bead); however, such a dependency is assumed throughout.

For bonds of length $d$ a temperature-shifted harmonic prior term was employed:
\begin{equation}
    W^{\text{bonds}}(d, \beta) = k(d-d_0)^2 + \frac{k'}{\beta} + V_0,
\end{equation}
where $\beta = \frac{1}{k_B T}$, $k$ and $d_0$ are the force constant and the equilibrium distance, respectively, $k'$ regulates the temperature scaling, and $V_0$ is a constant offset. While this is an approximation, we found it effective for fitting the system data within the target temperature regime.

An angular prior term was considered for angles $\theta$ between three consecutive CG beads, using the functional form:
\begin{equation}
    W^{\text{angles}}(\theta, \beta) = \sum_{i=1}^4 \left(a_i + \frac{b_i}{\beta}\right)\cos^i(\theta) + \frac{c}{\sin^2(\theta)} + V_0,
\end{equation}
where $a_i$ and $b_i$ linearly modulate the temperature dependence of the $i$-th term of the expansion, $c$ regulates the magnitude of the restriction introduced by the inverse squared sine term, and $V_0$ is a constant offset. This functional form is useful for fitting double-well potentials in $\cos(\theta)$ while adding a potential barrier at $\theta \in \{0, \pi\}$ via the $\sim 1/\sin^2(\theta)$ term. This restricted angular potential prevents singularities in the definition of the torsional potential and avoids numerical instabilities during MD simulations~\cite{bulacu2013improved}.

Dihedral priors were modeled using a temperature-shifted truncated Fourier expansion:
\begin{equation}
    W^{\text{dihedral}}(\psi, \beta) = \sum_{n=1}^{6} \left(\alpha_n\sin(n\psi) + \gamma_n \cos(n\psi)\right) + \frac{\delta}{\beta} + V_0,
\end{equation}
where $\alpha_n$ and $\gamma_n$ are the coefficients for each sinusoidal and cosinusoidal term in the expansion, $\delta$ regulates the temperature scaling, and $V_0$ is, again, a constant offset.

For the repulsion prior energy term, applied to all bead pairs separated by at least two bonds, we used the form:
\begin{equation}
    W^{\text{repulsion}}(d) = \left( \frac{\sigma}{d} \right)^6,
    \label{equation:repulsion_prior}
\end{equation}
where $\sigma$ represents the excluded volume associated with the CG beads, estimated for all bead pairs as the minimum pairwise distance observed in the reference atomistic MD at the highest temperature (\SI{380}{\kelvin}).

\subsection*{Temperature-Independent Models}
All temperature-independent models utilize analogous prior formulations where the temperature dependency is omitted (i.e., $k'=0$ for bond priors, $b_i=0$ for angular priors, and $\delta=0$ for dihedral priors). In these cases, the priors are fitted exclusively at the reference temperature.

\subsection*{Exact derivation of temperature dependency for harmonic potential}
If we consider a feature $x(\mathbf{R})$ function of the CG coordinates that enters the energy function in an approximately quadratic form, e.g. the distance for bonded beads, its Boltzmann distribution can be approximated with a Gaussian: 
\[
    \mathbb{P}(x)=\frac{e^{-\beta k(x-x_0)^2}}{\mathcal{Z}} \qquad \text{where}\quad \mathcal{Z} = \sqrt{\frac{\pi}{k\beta}}\sim\sqrt{T}
\]
where $k$ and $x_0$ are scalar constants. If we now consider the free energy associated with this distribution, we have 
\[
    A(x)=-\frac{1}{\beta}\ln\mathbb{P}(x) = k(x-x_0)^2 + \frac{1}{2}\kb T(\ln T+C)
\]
where the last therm can be typically approximated as linear, yielding a candidate function for fitting via Boltzmann inversion 
\[
    A(x)\simeq k(x-x_0)^2 + a T + A_0.
\]


\section{Multi temperature dataset}
To ensure a thorough sampling of the conformational landscape across various thermal regimes, we performed simulations at five distinct temperatures: \SI{300}{\kelvin}, \SI{320}{\kelvin}, \SI{350}{\kelvin}, \SI{380}{\kelvin}, and \SI{400}{\kelvin}. 
Starting configurations were obtained by first performing a single, extensive MD simulation at \SI{400}{\kelvin} to sample a broad region of the phase space. 
This trajectory was projected onto a two-dimensional Time-lagged Independent Component (TIC)~\cite{perez2013identification} space, where a regular space clustering was performed to select 1,000 representative structures. 
These same 1,000 configurations were then used as the starting points for sets of shorter, independent MD trajectories at each of the five target temperatures, ensuring a diverse and consistent ensemble of starting states and transition paths across the studied thermal range.
All simulations were conducted using the \texttt{OpenMM} engine~\cite{eastman2023openmm}, employing a Langevin integrator with a friction coefficient of \SI{1}{\per\pico\second}. The atomistic interactions were described by the AMBER99SB-ILDN protein force field~\cite{lindorff2010improved} in conjunction with the TIP3P explicit water model \cite{jorgensen1983comparison}.The initial structures for these simulations were sampled from a preliminary high-temperature trajectory to avoid starting-frame bias. Each independent trajectory reached a length of 50 ns, with the initial \SI{1}{\nano\second} discarded as equilibration. This results in an aggregate sampling time of \SI{50}{\micro\second} per temperature, totaling \SI{250}{\micro\second} of MD data for the entire dataset. System coordinates, forces, and relevant thermodynamic observables including the potential energy, were recorded every 2 ps. The resulting data is organized as a collection of indexed trajectories containing all-atom configurations, their corresponding coarse-grained projections (based on $C_\alpha$ mapping), and associated thermodynamic quantities. A comprehensive summary of the stored data fields is provided in Table~\ref{tab:avail_properties}.

\begin{table}[ht]
\centering
\caption{\textbf{Available multi temperature dataset properties:} Overview of all physical properties provided in the multi-temperature Chignolin dataset. For each entry, we report the corresponding units, array dimensions, and the specific key used to access the data within the \textit{hdf5} storage file.}
\label{tab:avail_properties}
    \begin{tabular}{p{3.4cm}ccr}
    \toprule
    \textbf{Property} & \textbf{Unit} & \textbf{Dimensions} &\textbf{HDF5 keys} \\
    \midrule
    Atomic positions & \AA & 3n & \texttt{coordinates} \\[1.5ex]
    Atomic forces & kcal/mol/\AA & 3n & \texttt{forces} \\[1.5ex]
    Coarse grained positions & \AA & 3N & \texttt{coordinates\_cg} \\[1.5ex]
    Coarse grained basic forces & kcal/mol/\AA & 3N & \texttt{cg\_basic\_forces} \\[1.5ex]
    Coarse grained optimized forces & kcal/mol/\AA & 3N & \texttt{cg\_optim\_forces} \\[1.5ex]
    Potential energy of the protein in vacuum & kcal/mol & 1 & \texttt{protein\_potential\_energy} \\[1.5ex]
    Potential energy of the CA atoms in vacuum& kcal/mol & 1 & \texttt{ca\_potential\_energy} \\[1.5ex]
    Total kinetic energy & kcal/mol & 1 & \texttt{total\_kinetic\_energy} \\[1.5ex]
    Total potential energy & kcal/mol & 1 & \texttt{total\_potential\_energy} \\[1.5ex]
    Density of the system & g/ml & 1 & \texttt{total\_density}\\
    \bottomrule
    \end{tabular}
\end{table}

\section{MSM re-weighting of reference simulations and $\Delta G$ calculation}

To obtain converged reference free energy profiles from the short atomistic simulations at each temperature, we estimated a Markov State Model (MSM)~\cite{prinz2011markov} for every temperature set and used it to re-weight the trajectory samples.

Initially, we performed a Time-lagged Independent Component Analysis (TICA)~\cite{perez2013identification} on the trajectories, utilizing all pairwise distances between the protein $C_\alpha$ atoms as input features. The MSMs were then constructed by projecting the system onto the first two dominant TICs. To discretize the conformational space into microstates, we performed $k$-means clustering with 200 centers for the AA reference data. The transition matrix was subsequently estimated using a maximum-likelihood procedure.

The MSM lag time was set to \SI{16}{\nano\second} for simulations at \SI{380}{\kelvin} and \SI{400}{\kelvin}, and \SI{20}{\nano\second} for \SI{350}{\kelvin} and \SI{320}{\kelvin} and \SI{24}{\nano\second} for \SI{300}{\kelvin}. These lag times were determined by monitoring the convergence of the re-weighted free energy as a function of increasing lag time. Specifically, we assessed convergence by computing the Jensen-Shannon (JS) divergence between the Free Energy Surfaces (FES) at successive lag times and by evaluating the Mean Absolute Error (MAE) of the MSM-derived weights.

To compute the $\Delta G$ values, we used the same method described in ref.~\cite{zaporozhets2023multibody}: first a Perron Cluster Cluster Analysis (PCCA+)~\cite{roblitz2013fuzzy} was performed to assign to each microstate the probability of belonging to a particular cluster. The number of clusters was set to 3 for all models, except the atomistic \SI{380}{\kelvin} were 4 clusters were used. From this fuzzy assignment, a hard one was obtained introducing a membership cutoff. For a given cutoff value, a corresponding $\Delta G$ is defined as 
\begin{equation}
    \Delta G = -k_B T \ln \left( \frac{\mathbb{P}_{\text{unfolded}}}{\mathbb{P}_{\text{folded}}} \right).
\end{equation}
The reported $\Delta G$ were estimated by considering the average of $\Delta G$ values obtained from 454 equally spaced membership cutoff values between 0.5 and 0.95 and considering as error their standard deviation. All the numerical results are reported in Table~\ref{tab:deltag}.

\begin{table}[h]
    \centering
    \caption{Estimated $\Delta G$ values for different models at different temperatures.}
    \label{tab:deltag}
    \setlength{\tabcolsep}{8pt}
    \begin{tabular}{lccccc}
        \toprule
         Model & $\Delta G$ \SI{300}{\kelvin} & $\Delta G$ \SI{320}{\kelvin} & $\Delta G$ \SI{350}{\kelvin} & $\Delta G$ \SI{380}{\kelvin} & $\Delta G$ \SI{400}{\kelvin}\\
               & kcal/mol & kcal/mol & kcal/mol & kcal/mol & kcal/mol\\
         \midrule
         Reference & $1.80\pm0.56$ & $1.59\pm0.24$ & $1.49\pm0.48$ & $1.41\pm0.77$ & $1.02\pm0.77$ \\
         $T$-model & $2.34\pm0.30$ & $2.13\pm0.26$ & $1.56\pm0.26$ & $0.79\pm0.16$ & $0.27\pm0.21$ \\
         320 model & $2.55\pm0.18$ & $2.49\pm0.48$ & $0.70\pm0.16$ & $-0.63\pm0.06$ & $-1.52\pm0.03$ \\
         380 model & $3.04\pm0.01$ & $2.78\pm0.09$ & $2.10\pm0.23$ & $0.87\pm0.18$ & $0.07\pm0.15$ \\
         \bottomrule
    \end{tabular}
\end{table}

The MSMs for the neural network-based coarse-grained (CG) models were obtained following the same procedure described for the atomistic MSMs, with the exception that 100 cluster centers were used for the $k$-means discretization. A lag time of 160 CG MD steps was employed for all CG MSMs. The statistical convergence of all models was assessed by verifying the plateau of the implied timescales as a function of the lag time.

All MSM and TICA analyses were performed using the \texttt{Deeptime} library~\cite{hoffmann2021deeptime}.

\section{Heat capacity calculation}

The isochoric heat capacity $c_V$ is obtained by evaluating the temperature derivative of the mean potential energy:
\[
    c_V = \frac{\partial}{\partial T} \langle u \rangle = \frac{\partial}{\partial T} \langle U_W \rangle.
\]
Because our dataset consists of a discrete set of temperatures, a naive finite-difference approximation of this derivative can be highly sensitive to statistical noise and lead to large fluctuations. To overcome this source of numerical instability, we fit a third-order spline to the predicted mean potential energies across the simulated temperatures and analytically evaluate the derivative of the spline to obtain $c_V$.

To estimate the statistical uncertainty associated with the heat capacity, we adopted a hierarchical resampling strategy based on block-averaging. First, to account for temporal correlations within the time series, each potential energy trajectory was partitioned into contiguous, non-overlapping blocks of length $B$. The correlated raw frames within each block were then replaced by their arithmetic mean, reducing each original trajectory to a shorter sequence of uncorrelated, block-averaged energy values. Finally, a standard bootstrap procedure was performed globally across this reduced dataset, resampling both the block-averaged frames within the trajectories and the independent trajectories themselves to compute the standard error of $c_V$.

\begin{figure}
    \centering
    \includegraphics[width=1.\linewidth]{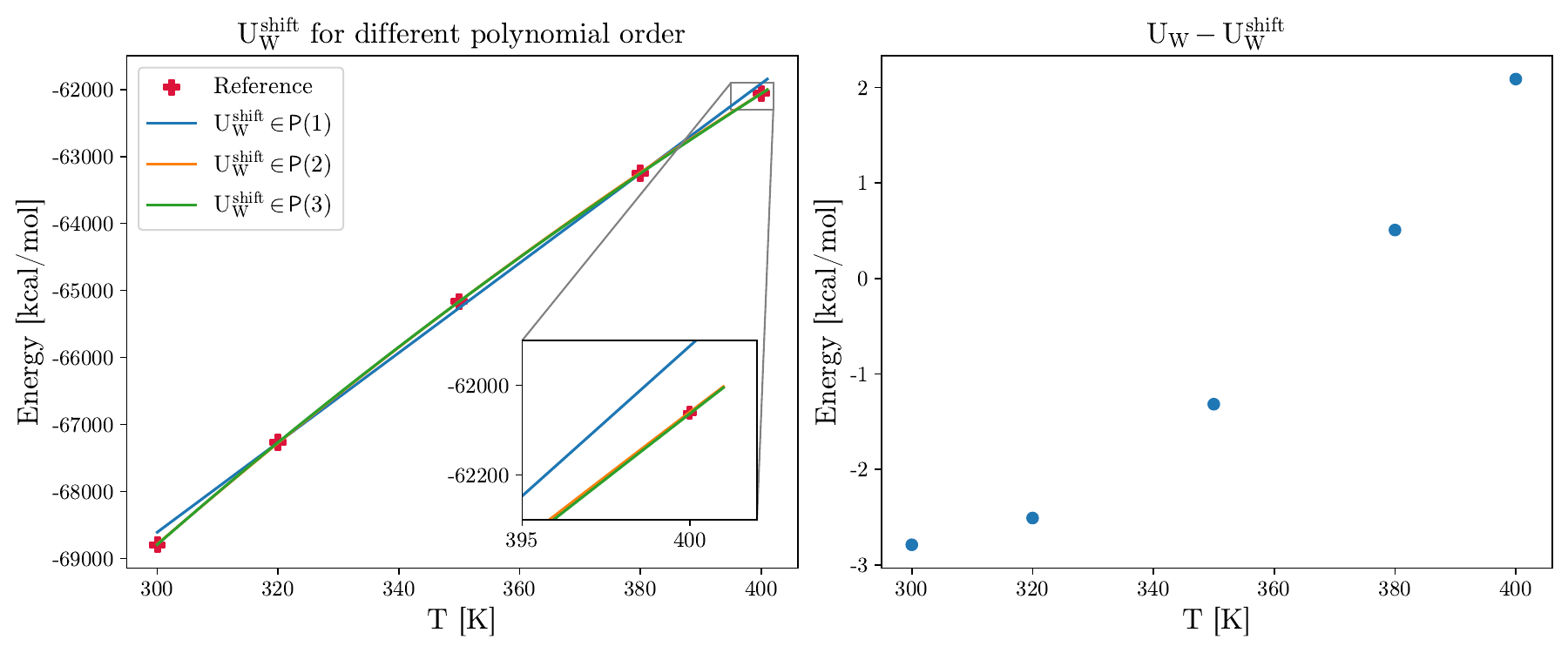}
    \caption{\textbf{Contributions of different terms to $U_W$.} Left: Polynomial fits of order $n \in \{1, 2, 3\}$ for the temperature-dependent shift component, $U_W^{\text{shift}}(T)$. Right: Mean potential energy predicted by the neural network potential, calculated excluding the global temperature shift. Error bars are smaller than the marker size and are omitted for visual clarity.}
    \label{fig:u_W_contributions}
\end{figure}

To identify an appropriate $B$ value for the internal bootstrap loop, we analyzed the convergence of the standard error of $c_V$ as a function of $B$, choosing a value where the error estimate stabilizes. For the coarse-grained (CG) simulations, we utilized a block size of $5 \times 10^3$ frames. For the reference atomistic (AA) trajectories, which were strided by a factor of 10 to yield a \SI{20}{\pico\second} sampling interval, we selected a block size of $6 \times 10^2$ frames.

\begin{table}[h]
    \centering
    \small
    \caption{Computed $c_V$ values for reference and temperature dependent model using different polynomial order for fitting the correction $U_W^{\text{shift}}$ for different temperatures.}
    \label{tab:cv}
    \setlength{\tabcolsep}{8pt}
    \begin{tabular}{lccccc}
        \toprule
         Model & $c_V$ \SI{300}{\kelvin} & $c_V$ \SI{320}{\kelvin} & $c_V$ \SI{350}{\kelvin} & $c_V$ \SI{380}{\kelvin} & $c_V$ \SI{400}{\kelvin}\\
               & kcal/mol/K & kcal/mol/K & kcal/mol/K & kcal/mol/K & kcal/mol/K\\
         \midrule
         Reference & $79.129 \pm 0.021$ & $73.658 \pm 0.006$ & $66.751 \pm 0.006$ & $61.127 \pm 0.007$ & $57.938 \pm 0.025$ \\
         $T$-model-$\mathcal{P}(1)$ & $66.959 \pm 0.008$ & $66.984 \pm 0.003$ & $67.009 \pm 0.002$ & $67.029 \pm 0.003$ & $67.045 \pm 0.013$ \\
         $T$-model-$\mathcal{P}(2)$ & $77.622 \pm 0.008$ & $73.506 \pm 0.003$ & $67.319 \pm 0.002$ & $61.128 \pm 0.003$ & $57.003 \pm 0.013$ \\
         $T$-model-$\mathcal{P}(3)$ & $79.015 \pm 0.008$ & $73.685 \pm 0.003$ & $66.751 \pm 0.002$ & $61.101 \pm 0.003$ & $58.052 \pm 0.013$ \\
         \bottomrule
    \end{tabular}
\end{table}

\section{Model details}

The neural network models were implemented using the CGSchNet architecture~\cite{husic2020coarse, charron2025navigating}. 
Coarse-grained beads are mapped to an index corresponding to their residue identity; from this, they are transformed into initial node features $\mathbf{h}^0_i$ via a learnable embedding layer. These features are iteratively refined through $T$ interaction blocks:
\begin{equation}
    \mathbf{h}^t_i = \mathbf{h}^{t-1}_i + \mathbf{\Theta}^{t,3}\sigma \left( \mathbf{\Theta}^{t,2} \sum_{j\in\mathcal{N(i)}} \mathbf{\Theta}^{t,1}\mathbf{h}^{t-1}_j \odot \text{MLP}(E_{ij})\, \mathcal{C}(r_{ij}) \right),
\end{equation}
where $E_{ij}$ represents the inter-bead distances expanded in a radial basis, the MLP acts as a filter-generating network, $\left\{\Theta^{t,l}\right\}$ are learnable matrices,  and $\mathcal{C}$ is a cosine cutoff function. 

Following the final node embedding update, $\mathbf{h}^T_i$ is transformed by a temperature embedding:
\begin{equation}
    \mathbf{h}^T_i = \mathbf{h}^T_i + \mathbf{h}^T_i \odot \boldsymbol{\theta}_{\beta} \left( \frac{\beta-\beta_{\text{min}}}{\beta_{\text{max}}-\beta_{\text{min}}} \right),
\end{equation}
where $\boldsymbol{\theta}_{\beta}$ is a learnable parameter vector, and $\beta_{\text{min}}$ and $\beta_{\text{max}}$ are the minimum and maximum inverse temperatures used for normalization. This temperature embedding is used only in the temperature-dependent model. 

The final Potential of Mean Force (PMF), denoted as $W$, is computed as a sum of local contributions via a readout MLP:
\begin{equation}
    W = \sum^{N}_{i=1} \text{MLP}(\mathbf{h}^T_i).
    \label{eq:readout}
\end{equation}
The CG forces are subsequently derived through the negative gradient of $W$ with respect to the CG coordinates $\mathbf{R}$. The CG internal energy and entropy contributions are obtained by applying the thermodynamic transformations defined in the main text to the network output.

\begin{table}[ht]
\centering
\caption{Hyperparameters used for the SchNet-based neural network models. Parameters are consistent across both temperature-dependent and temperature-independent architectures.}
\label{table:base_hyperparams}
\renewcommand{\arraystretch}{1.2} 
\begin{tabular}{lc} 
\toprule
\textbf{Hyperparameter} & \textbf{Value} \\
\midrule
Embedding Feature Size              & 128 \\ 
Number of Basis Functions           & 64 \\
Number of Filters                   & 128 \\
Interaction Blocks                  & 2 \\ 
Cutoff                              & 30.0 \AA \\
Distance Expansion Basis            & PhysNet basis~\cite{unke2019physnet} \\
Readout MLP                         & MLP (2 layers, [128, 64] neurons) \\
Activation Function                 & Tanh \\
Scaling $(\beta_{min},\beta_{max})$ & (0,2) \\
\bottomrule
\end{tabular}
\end{table}

\section{Model training}
All neural network (NN) models were trained using the \texttt{mlcg} package~\cite{charron2025navigating} based on the \texttt{PyTorch Lightning}~\cite{Falcon_PyTorch_Lightning_2019} framework. Optimization was performed using the AdamW~\cite{loshchilov2017decoupled} optimizer with a learning rate of $10^{-4}$ and a weight decay of 0.01, employing a batch size of 512 molecules. The loss function was defined as a weighted sum of the Mean Squared Error (MSE) for both forces and coarse-grained energies, as detailed in the main text. Specifically, the weights were set to 1.0 for the force contribution and 0.005 for the energy contribution.

Each model was trained for at least 250 epochs. Following the observation in previous studies that the minimum validation loss is not a sufficient metric to identify the optimal model for MD simulations~\cite{fu202h2forces, charron2025navigating,durumeric2026learning}, we adopted a selection strategy based on sampling performance. Starting from epoch 80, models were saved and simulated every 20 epochs to evaluate their stability and accuracy. The final models were selected at epoch 240 for the temperature-dependent model, epoch 120 for the model trained at \SI{380}{\kelvin}, and epoch 240 for the model trained at \SI{320}{\kelvin}, as these demonstrated the best matching of their respective reference free energy.

\section{Coarse-Grained Molecular Dynamics Simulations}
Coarse-grained (CG) Molecular Dynamics simulations were performed using the \texttt{mlcg} package, employing a Langevin integrator with the BAOA(F)B scheme~\cite{charron2025navigating}. Every model was used to simulate the system at five distinct temperatures: \SI{300}{\kelvin}, \SI{320}{\kelvin}, \SI{350}{\kelvin}, \SI{380}{\kelvin}, and \SI{400}{\kelvin}. 

For each temperature, 50 independent trajectories were run in parallel on a single GPU, starting from a diverse set of initial configurations. We utilized an integration timestep of \SI{4}{\femto\second} and a friction coefficient of \SI{1}{\pico\second\tothe{-1}}. Each trajectory was evolved for $2 \times 10^6$ steps, with snapshots saved every 10 steps for subsequent analysis. The initial 100,000 steps of each simulation were discarded as equilibration. For each model, this protocol yielded a total of 250 independent trajectories across the temperature range, providing sufficient statistics for the MSM analysis and free energy calculations.

\clearpage


\begin{thebibliography}{10}

\bibitem{clementi2008coarse}
C.~Clementi.
\newblock Coarse-grained models of protein folding: toy models or predictive
  tools?
\newblock {\em Curr. Opin. Struc. Biol.}, 18(1):10--15, 2008.

\bibitem{stevens2023molecular}
J.~A. Stevens, F.~Gr{\"u}newald, P.~van Tilburg, et~al.
\newblock Molecular dynamics simulation of an entire cell.
\newblock {\em Front. Chem.}, 11:1106495, 2023.

\bibitem{charron2025navigating}
N.~E. Charron, K.~Bonneau, A.~S. Pasos-Trejo, et~al.
\newblock Navigating protein landscapes with a machine-learned transferable
  coarse-grained model.
\newblock {\em Nat. Chem.}, 17:1284–1292, 2025.

\bibitem{noid2024rigorous}
W.~G. Noid, R.~J. Szukalo, K.~M. Kidder, and M.~C. Lesniewski.
\newblock Rigorous progress in coarse-graining.
\newblock {\em Annu. Rev. Phys. Chem.}, 75(1):21--45, 2024.

\bibitem{noid2008multiscale}
W.~G. Noid, J.~Chu, G.~S. Ayton, et~al.
\newblock The multiscale coarse-graining method. i. a rigorous bridge between
  atomistic and coarse-grained models.
\newblock {\em J. Chem. Phys.}, 128(24), 2008.

\bibitem{noid2013perspective}
W.~G. Noid.
\newblock Perspective: Coarse-grained models for biomolecular systems.
\newblock {\em J. Chem. Phys.}, 139(9), 2013.

\bibitem{jin2022bottom}
J.~Jin, A.~J. Pak, A.~E.~P. Durumeric, et~al.
\newblock Bottom-up coarse-graining: Principles and perspectives.
\newblock {\em J. Chem. Theory Comput.}, 18(10):5759--5791, 2022.

\bibitem{shell2008relative}
M.~S. Shell.
\newblock The relative entropy is fundamental to multiscale and inverse
  thermodynamic problems.
\newblock {\em J. Chem. Phys.}, 129(14), 2008.

\bibitem{wang2019machine}
J.~Wang, S.~Olsson, C.~Wehmeyer, et~al.
\newblock Machine learning of coarse-grained molecular dynamics force fields.
\newblock {\em ACS Cent. Sci.}, 5(5):755--767, 2019.

\bibitem{durumeric2023machine}
A.~E.~P. Durumeric, N.~E. Charron, C.~Templeton, et~al.
\newblock Machine learned coarse-grained protein force-fields: Are we there
  yet?
\newblock {\em Curr. Opin. Struc. Biol.}, 79:102533, 2023.

\bibitem{husic2020coarse}
B.~E. Husic, N.~E. Charron, D.~Lemm, et~al.
\newblock Coarse graining molecular dynamics with graph neural networks.
\newblock {\em J. Chem. Phys.}, 153(19), 2020.

\bibitem{majewski2023machine}
M.~Majewski, A.~P{\'e}rez, P.~Th{\"o}lke, et~al.
\newblock Machine learning coarse-grained potentials of protein thermodynamics.
\newblock {\em Nat. Commun.}, 14(1):5739, 2023.

\bibitem{durumeric2026learning}
A.~E.~P. Durumeric, Y.~Chen, A.~S. Pasos-Trejo, et~al.
\newblock Learning data-efficient coarse-grained molecular dynamics from forces
  and noise.
\newblock {\em Nat. Commun.}, 2026.

\bibitem{wang2021multi}
J.~Wang, N.~E. Charron, B.~Husic, et~al.
\newblock Multi-body effects in a coarse-grained protein force field.
\newblock {\em J. Chem. Phys.}, 154(16), 2021.

\bibitem{dunn2016van}
N.~J.~H. Dunn, T.~T. Foley, and W.~G. Noid.
\newblock Van der waals perspective on coarse-graining: Progress toward solving
  representability and transferability problems.
\newblock {\em Acc. Chem. Res.}, 49(12):2832--2840, 2016.

\bibitem{kidder2021energetic}
K.~M. Kidder, R.~J. Szukalo, and W.~G. Noid.
\newblock Energetic and entropic considerations for coarse-graining.
\newblock {\em Eur. Phys. J. B}, 94(7):153, 2021.

\bibitem{szukalo2021investigating}
R.~J. Szukalo and W.~G. Noid.
\newblock Investigating the energetic and entropic components of effective
  potentials across a glass transition.
\newblock {\em J. Phys. Condens. Matter}, 33(15):154004, 2021.

\bibitem{mussi2025predicting}
L.~M. Mussi and W.~G. Noid.
\newblock Predicting energetic and entropic driving forces with coarse-grained
  models.
\newblock {\em J. Chem. Phys.}, 163(8), 2025.

\bibitem{kramer2023statistically}
A.~Kr{\"a}mer, A.~E.~P. Durumeric, N.~E. Charron, et~al.
\newblock Statistically optimal force aggregation for coarse-graining molecular
  dynamics.
\newblock {\em J. Phys. Chem. Lett.}, 14(17):3970--3979, 2023.

\bibitem{hendriks2020linearly}
J.~Hendriks, C.~Jidling, A.~Wills, and T.~Sch{\"o}n.
\newblock Linearly constrained neural networks.
\newblock {\em arXiv preprint arXiv:2002.01600}, 2020.

\bibitem{eastman2023openmm}
P.~Eastman, R.~Galvelis, R.~P. Pel{\'a}ez, et~al.
\newblock Openmm 8: molecular dynamics simulation with machine learning
  potentials.
\newblock {\em J. Phys. Chem.~B}, 128(1):109--116, 2023.

\bibitem{ruza2020temperature}
J.~Ruza, W.~Wang, D.~Schwalbe-Koda, et~al.
\newblock Temperature-transferable coarse-graining of ionic liquids with dual
  graph convolutional neural networks.
\newblock {\em J. Chem. Phys.}, 153(16), 2020.

\bibitem{shinkle2024thermodynamic}
E.~Shinkle, A.~Pachalieva, R.~Bahl, et~al.
\newblock Thermodynamic transferability in coarse-grained force fields using
  graph neural networks.
\newblock {\em J. Chem. Theory Comput.}, 20(23):10524--10539, 2024.

\bibitem{perez2013identification}
G.~P{\'e}rez-Hern{\'a}ndez, F.~Paul, T.~Giorgino, et~al.
\newblock Identification of slow molecular order parameters for markov model
  construction.
\newblock {\em J. Chem. Phys.}, 139(1), 2013.

\bibitem{prinz2011markov}
J.-H. Prinz, H.~Wu, M.~Sarich, et~al.
\newblock Markov models of molecular kinetics: Generation and validation.
\newblock {\em J. Chem. Phys.}, 134(17), 2011.

\bibitem{husic2018markov}
B.~E. Husic and V.~S. Pande.
\newblock Markov state models: From an art to a science.
\newblock {\em J. Am. Chem. Soc.}, 140(7):2386--2396, 2018.

\bibitem{hoffmann2021deeptime}
M.~Hoffmann, M.~K. Scherer, T.~Hempel, et~al.
\newblock Deeptime: a python library for machine learning dynamical models from
  time series data.
\newblock {\em Mach. Learn.: Sci. Technol.}, 2021.

\bibitem{thaler2022deep}
S.~Thaler, M.~Stupp, and J.~Zavadlav.
\newblock Deep coarse-grained potentials via relative entropy minimization.
\newblock {\em J. Chem. Phys.}, 157(24), 2022.

\bibitem{roblitz2013fuzzy}
S.~R{\"o}blitz and M.~Weber.
\newblock Fuzzy spectral clustering by pcca+: application to markov state
  models and data classification.
\newblock {\em Adv. Data Anal. Classif.}, 7(2):147--179, 2013.

\bibitem{bulacu2013improved}
M.~Bulacu, N.~Goga, W.~Zhao, et~al.
\newblock Improved angle potentials for coarse-grained molecular dynamics
  simulations.
\newblock {\em J. Chem. Theory Comput.}, 9(8):3282--3292, 2013.

\bibitem{lindorff2010improved}
K.~Lindorff-Larsen, S.~Piana, K.~Palmo, et~al.
\newblock Improved side-chain torsion potentials for the amber ff99sb protein
  force field.
\newblock {\em Proteins}, 78(8):1950--1958, 2010.

\bibitem{jorgensen1983comparison}
W.~L. Jorgensen, J.~Chandrasekhar, J.~D. Madura, et~al.
\newblock Comparison of simple potential functions for simulating liquid water.
\newblock {\em J. Chem. Phys.}, 79(2):926--935, 1983.

\bibitem{zaporozhets2023multibody}
I.~Zaporozhets and C.~Clementi.
\newblock Multibody terms in protein coarse-grained models: A top-down
  perspective.
\newblock {\em J. Phys. Chem.~B}, 127(31):6920--6927, 2023.

\bibitem{unke2019physnet}
O.~T. Unke and M.~Meuwly.
\newblock Physnet: A neural network for predicting energies, forces, dipole
  moments, and partial charges.
\newblock {\em J. Chem. Theory Comput.}, 15(6):3678--3693, 2019.

\bibitem{Falcon_PyTorch_Lightning_2019}
W.~Falcon and {The PyTorch Lightning team}.
\newblock {PyTorch Lightning}, 2019.

\bibitem{loshchilov2017decoupled}
I.~Loshchilov and F.~Hutter.
\newblock Decoupled weight decay regularization.
\newblock {\em arXiv preprint arXiv:1711.05101}, 2017.

\bibitem{fu202h2forces}
X.~Fu, Z.~Wu, W.~Wang, et~al.
\newblock Forces are not enough: Benchmark and critical evaluation for machine
  learning force fields with molecular simulations.
\newblock {\em arXiv preprint arXiv:2210.07237}, 2022.

\end{thebibliography}

\end{document}